\documentclass[prl,aps,showpacs,twocolumn,superscriptaddress]{revtex4-1}
\usepackage{graphicx}
\usepackage{bm}
\usepackage{amsfonts}
\usepackage{amsmath}
\usepackage{amssymb}
\usepackage{color}
\usepackage{diagbox}
\usepackage{soul}

\newcommand{\St}{\mbox{\textit{St}}}
\newcommand{\Sts}{St_{\rm s}}

\newcommand{\K}{\mathrm{K}}
\newcommand{\A}{\mathrm{A}}

\newcommand{\ictsaddress}{International Centre for Theoretical Sciences, Tata Institute of Fundamental Research,
 Bangalore 560089, India}
\newcommand{\iiscaddress}{Department of Mathematics, Indian Institute of Science, Bangalore 560012, India}

\newcommand{\uiucaddress}{Department of Mechanical Science and Engineering, University of Illinois, at Urbana-Champaign, Illinois 61801}

\begin{document}
\title{Inertial spheroids in homogeneous, isotropic turbulence}
\author{Amal Roy}\email{amalchettisseril@gmail.com}
\affiliation{\iiscaddress}\affiliation{\ictsaddress}
\author{Anupam Gupta}\email{anupam1509@gmail.com}
\affiliation{\uiucaddress}
\author{Samriddhi Sankar Ray}\email{samriddhisankarray@gmail.com}
\affiliation{\ictsaddress}

\begin{abstract}

We study the rotational dynamics of {\it inertial} disks and rods in
three-dimensional, homogeneous isotropic turbulence. In particular, we show 
how the alignment and the decorrelation time-scales of such
spheroids depend, critically, on both the level of inertia and the aspect
ratio of these particles. These results illustrate the
effect of inertia---which leads to a preferential sampling of the local flow
geometry---on the statistics of both disks and rods in a turbulent flow. Our
results are important for a variety of natural and industrial settings where
the turbulent transport of asymmetric, spheroidal inertial particles is
ubiquitous. 

\end{abstract}

\maketitle

The dynamics of small, heavy inertial particles in a turbulent flow is at the
heart of several problems in statistical physics, fluid dynamics, astrophysics
and the atmospheric sciences. This is because particles advected by a flow are
ubiquitous in nature, industry and the laboratory. Typically, for particles
smaller than the Kolmogorov scale $\eta$ of the three-dimensional (3D) carrier
(turbulent) flow, the fluid-particle interaction is modeled as a one-way
coupling via the linear Stokes drag model~\cite{MR,Bec}.  This model, despite
its many simplifications, has been shown, over the years, to effectively mimic
the turbulent transport of small spherical particles (see, e.g.,
Ref.~\cite{sawPoF}). In the last few years a significant part of the theoretical and numerical 
studies of such problems has been carried out with an eye on the problem of  
spherical water droplets in warm clouds~\cite{sling,caustic,mehlig,bec2014,bec2016}. 

The spherical particle approach, though valid in many circumstances,
nevertheless fails when dealing with a wide class of transport problems where
it is known that the particulate matter is rod-like or disk-like. These
range from the motion of microorganisms~\cite{micro,rayEPL} to ice
crystals in clouds~\cite{ice}. Unlike the spherical case, such particles have
an added degree of freedom which, based on their geometry of the surrounding
flow, allows such non-spherical particles, henceforth called {\it
spheroids} to {\it rotate}, {\it spin}, and {\it tumble}. Broadly speaking, in
a dilute suspension, the advecting fluid velocity gradient tensor along its
trajectory determines the rotational dynamics of a given spheroid.  In recent
years there have been a lot of effort to understand the various aspects of the
dynamics of spheroids in both homogeneous, isotropic turbulence as well as in
channel flows.  Indeed it is known that such particles have complex dynamics not
only in turbulent flows but in simpler flow
configurations~\cite{simple} as well. Unfortunately the experimental
measurements have been by and large restricted to two-dimensional
flows~\cite{2D-exp} with only recent time-resolved measurements in
three-dimensional turbulence~\cite{vothPRL2012}. 

Studies of spheroids with inertia have largely been confined to the area of
turbulent channel flows~\cite{turb-channel,mortensenPoF08} with an emphasis on clustering and
turbophoresis. Even the fewer number of studies within the framework of homogeneous,
isotropic turbulence have tended to focus on the effect of gravity in the
settling of such spheroids~\cite{turb-homo,gustav,jucha} or limited to the effect of such
particles on turbulent modulation ~\cite{turb-modu}.  The issue of orientation
dynamics and the alignment of inertial spheroids along specific flow directions have
largely been an unexplored regime; it is important to note that aspects of 
this problem have been investigated for non-spherical \textit{tracers}
in turbulence (triaxial ellipsoids)~\cite{chevillardjfm} and perturbatively in the Kubo 
number for random flows~\cite{mehligPRL}.

Theoretically, there have been studies which have looked at the orientation
dynamics of rod-like particles in the absence of inertia, i.e., rods which
display a tracer-like behavior~\cite{wilkinson-NJP,guptaPRE14}. However in most cases of
turbulent transport these asymmetrical particles are inertial. In other words a
more complete description of the rotational dynamics of such particles need to
take into account the fact that such particles relax to the flow velocity not
instantaneously (as a tracer would) but with a finite time-lag, the so-called
Stokes time $\tau_p$. Furthermore if $\alpha$, which is a measure of the ratio
of the major and minor axes of the spheroid, denotes the degree and nature of
the spheroid (with $\alpha = 1$, a sphere; $\alpha \ll 1$, an oblate; and
$\alpha \gg 1$, a rod), the dynamics should depend not only on the Stokes number
$St = \tau_p/\tau_\eta$ (where, $\tau_\eta$ is the characteristic fluid
small-scale Kolmogorov time to be defined later) but on $\alpha$ as well.

We address this question in a detailed and systematic manner in this Rapid Communication by
using extensive numerical simulations covering a wide range in $\alpha$ and
$St$ to explore the different regimes of particle alignment and orientations in
fully developed turbulence. By using ideas of inertial effects on spheroids~\cite{anderson-prl}, we thus complement and build 
on the work of Pumir and
Wilkinson~\cite{wilkinson-NJP} (and Parsa {\it et al.}~\cite{vothPRL2012}) who were the first to study this problem but
only in the case of inertia-less rods. 

\begin{figure*}
  \begin{center}
    \includegraphics[width=1.0\textwidth]{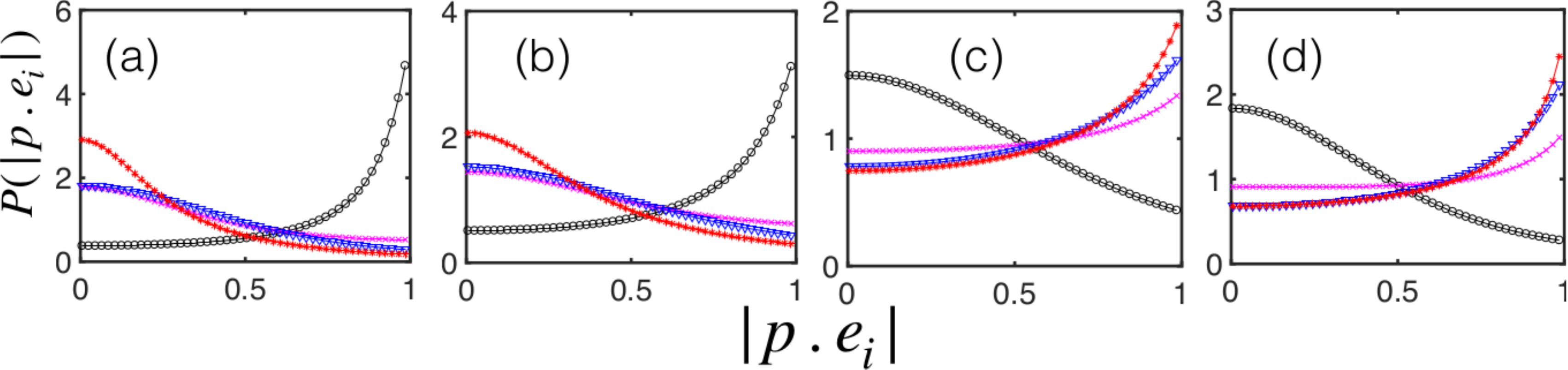}
    \vspace{-10pt}
    \caption{(color online) Representative plots of the probability density function of the 
    alignment of the orientation vector ${\bf p}$ with
    ${\bf e}_1$ (black circle $\circ$), ${\bf e}_2$ (pink cross $\times$), ${\bf e}_3$ (blue inverted triangle $\triangledown$) 
    and ${\bf e}_w$ (red asterisk $\ast$) for (a) $\alpha = 0.1$ (oblate), 
    (b) $\alpha = 0.5$ (c) $\alpha = 1.5$, and (d) $\alpha = 2.0$ (rod). These measurements are made for 
    particles with a  bare Stokes numbers $St_{\rm s} = 1.0$.  (In Table~\ref{table:Stokes}, the actual Stokes numbers 
    corresponding to the different particles are listed.)} 
  \label{fig:alignment}
  \end{center}
\end{figure*}

We begin by considering a spheroid of density $\rho_p$, with a symmetry axes of length $2c$ and the two 
equal axis of length $2a$, such that the ratio $\alpha = c/a$ characterizes the nature 
of the spheroid, moving with a velocity ${\bf v}$, and advected by a carrier fluid with 
velocity ${\bf u}$. In the most general case, the drag felt by a non-spherical particle is 
characterized by its resistance tensor $\K$~\cite{brenner} and the use of quaternion 
algebra in recent years~\cite{wachem} provides a 
convenient route to study the problem in its most general setting [see, e.g., Voth and Soldati~\cite{vothARFM} and references therein]. 
The equations of translational motion of the center of the spheroid ${\bf r}$ are given by the Stokes drag model:
\begin{equation}
\frac{d{\bf r}}{dt} = {\bf v}; \quad \frac{d{\bf v}}{dt} = -\frac{\A^{\mathrm T}\K \A}{6\pi a \alpha} \frac{[{\bf v} - {\bf u}]}{\tau_s},
\label{eq}
\end{equation}
where the carrier fluid velocity ${\bf u}$ above is evaluated at the particle position ${\bf r}$. 
The Stokes time for a spherical particle of radius $a$ is given by $\tau_s = 2\rho_pa^2/9\rho_f\nu$, 
where $\rho_f$ is the density and $\nu$ is the kinematic viscosity of the carrier flow.
The details of the resistence tensor $\K$ and the orthogonal transformation matrix $\A$ are described in Ref.~\cite{mortensenPoF08} for prolate spheroids and Ref.~\cite{challabotlajfm15} for oblate spheroids.
The Stokes time $\tau_p$, based on isotropic particle orientation and the inverse of the resistance tensor, differs from the more familiar spherical case $\tau_s$ to 
take into account the asymmetry of the particle~\cite{anderson-prl}, \begin{equation}
\tau_p = 
\begin{cases}
\tau_s \frac{\alpha\{\pi - 2\tan^{-1}[\alpha ({1 - \alpha^2})^{-1/2}]\}}{2\sqrt{1-\alpha^2}} \quad \alpha < 1 \\
\tau_s \frac{\alpha\ln[\alpha + \sqrt{\alpha^2-1}]}{\sqrt{\alpha^2 -1}}\quad \alpha > 1.
\end{cases}
\label{taup}
\end{equation} 
 We see immediately that for $\alpha = 1$, which corresponds to a spherical particle since 
$a = c$, the $\tau_p = \tau_s$ via the definition above by setting $\alpha = 1$. For convenience, we define a {\it bare} Stokes number 
$St_{\rm s} = \tau_s/\tau_\eta$; the actual Stokes number $St$ will of course depend on the value $\alpha$ via \eqref{taup}; in the
spherical case $St \equiv St_{\rm s}$. In Table~\ref{table:Stokes}, we list all the values of $\alpha$ and the Stokes numbers that 
we have used in our simulations.

For asymmetric particles $\alpha \neq 1$, along with the translational motion (defined above), the instantaneous orientation is 
vital to understand the full dynamics of such spheroids. Intuitively, the direction of the orientation vector ${\bf p}$ for a given 
spheroid, with a given $\tau_p$ and $\alpha$, is determined by the local flow geometry. For a given, generic complex flow, the local 
geometry is determined by the fluid-velocity-gradient tensor (traceless for incompressible flows), evaluated at the particle position 
${\mathcal A}_{ij} = \frac{\partial u_i}{\partial r_j}$.
It is useful to split this fluid-velocity-gradient tensor ${\bf \mathcal A} = {\bf S} + {\bf \Omega}$ 
into a symmetric part, the strain rate, ${\bf S}^T = {\bf S}$ and an antisymmetric part, the vorticity tensor, 
${\bf \Omega}^T = -{\bf \Omega}$. This decomposition is especially useful to write the equation for the orientation 
vector ${\bf p}$, the so-called Jeffery equation~\cite{jeffery}
\begin{equation}
\frac{d{\bf p}}{dt} = {\bf \Omega \, p} + \frac{\alpha^2 - 1}{\alpha^2 + 1} \left [{\bf S \, p} - ({\bf p \cdot S \, p})\,{\bf p} \right ], 
\label{jeffery}
\end{equation}
where the strain rate and vorticity tensor are instantaneous measurements at the (inertial) particle position.

It is important to stress that we are approximating the particle dynamics 
by ignoring the inertia associated with its rotational dynamics. Such a simplification is justified 
because it has been shown that the typical relaxation timescale associated with 
the rotational dynamics is an order of magnitude smaller than the $\tau_p$~\cite{anderson-prl,marchioli}.

\begin{figure*}
  \begin{center}
    \includegraphics[width=1.0\textwidth]{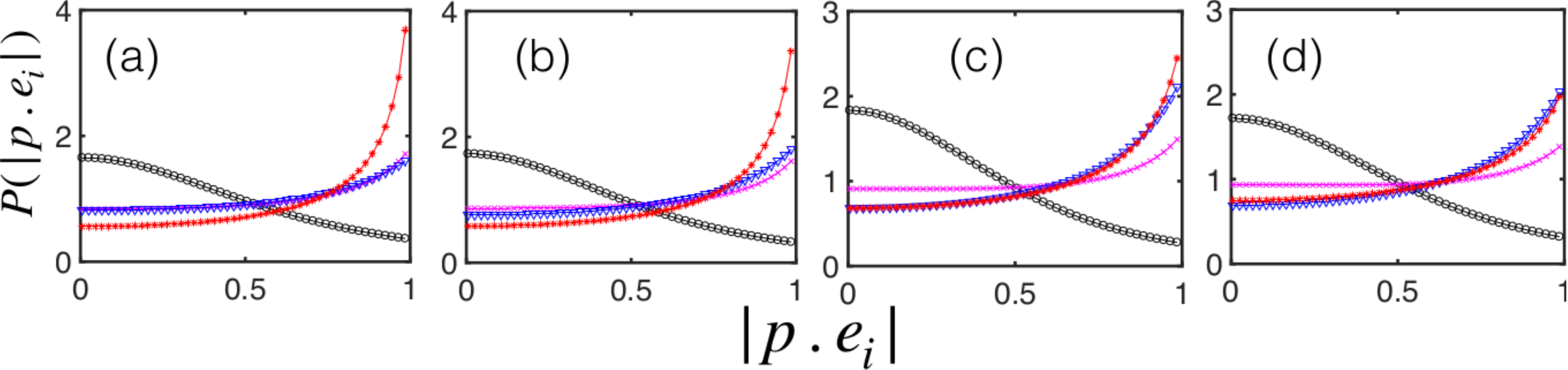}
    \vspace{-10pt}
      \caption{(color online) Representative plots of the probability density function of the 
     alignment of the orientation vector ${\bf p}$ of a rod ($\alpha = 2.0$) with
    ${\bf e}_1$ (black circle $\circ$), ${\bf e}_2$ (pink cross $\times$), ${\bf e}_3$ (blue inverted triangle $\triangledown$) 
    and ${\bf e}_w$ (red asterisk $\ast$) for (a) $St = 0.0$, (b) $St = 0.152$, (c) $St = 0.76$, and (d) $St = 4.56$.
} 
\label{fig:alignment_vs_St}
  \end{center}
\end{figure*}

\begin{table}
\begin{tabular}{|l||*{7}{c|}}\hline
\backslashbox{$St_{\rm s}$}{$\alpha$}
&\makebox[3em]{\bf 0.1}&\makebox[3em]{\bf 0.5}&\makebox[3em]{\bf 0.9}
&\makebox[3em]{\bf 1.0}&\makebox[3em]{\bf 1.1}&\makebox[3em]{\bf 1.5}&\makebox[3em]{\bf 2.0}\\\hline\hline
&\makebox[3em]{oblate}&\makebox[3em]{}&\makebox[3em]{}
&\makebox[3em]{sphere}&\makebox[3em]{}&\makebox[3em]{}&\makebox[3em]{rod}\\\hline\hline
{\bf 0.0} & 0.0 & 0.0 & 0.0 & 0.0 & 0.0 & 0.0 & 0.0\\\hline
{\bf 0.1} & 0.015 & 0.067 & 0.247 & 0.1 & 0.106 & 0.129 & 0.152\\\hline
{\bf 0.5} & 0.075 & 0.34 & 1.24 & 0.5 & 0.53 & 0.65 & 0.76 \\\hline
{\bf 1.0} & 0.15 & 0.67 & 2.47 & 1.0 &  1.064 & 1.29 & 1.52 \\\hline
{\bf 2.0} & 0.30 & 1.34 & 4.94 & 2.0 & 2.13 & 2.58 & 3.04 \\\hline
{\bf 3.0} & 0.45 & 2.01 & 7.41& 3.0 &  3.19 & 3.87 & 4.56 \\\hline
\end{tabular}
\caption{Values of the aspect ratios $\alpha$, the bare Stokes numbers $St_{\rm s}$, and the actual Stokes numbers $St$ 
for the different sets of particles that we have used in our simulations (see text).}
\label{table:Stokes}
\end{table}

We finally turn our attention to the advecting or carrier fluid velocity ${\bf u}$. Since we study the spheroid in a three-dimensional, incompressible 
turbulent flow, we obtain the velocity field as a solution of the forced three-dimensional Navier-Stokes equation :
\begin{equation}
\frac{\partial {\bf u}}{\partial t} + {\bf u}\cdot{\nabla}{\bf u} = \nu \nabla^2{\bf u} - \frac{\nabla P}{\rho_f} + {\bf f}, 
\label{NS}
\end{equation}
augmented by the incompressibility constraint $\nabla \cdot {\bf u} = 0$, 
where $P$ is the pressure and the forcing ${\bf f}$ drives the system to a statistically steady state. We recall that a three-dimensional turbulent flow
are characterised by the Kolmogorov micro-scales for length $\eta = \left (\frac{\nu^3}{\epsilon}\right )^{1/4}$, 
time $\tau_\eta = \left (\frac{\nu}{\epsilon}\right )^{1/2}$, and velocity $u_\eta = \left (\nu\epsilon \right )^{1/4}$. 
These definitions allow us in a unique way, which allows a comparison between experiments, 
numerical simulations, and theory, to define the Stokes number $St = \tau_p/\tau_\eta$. 
We should also note that our model, and hence the results, are valid only for $a, c \ll \eta$.

Before we discuss the various results, let us briefly outline the numerical strategy used in our calculations. (We refer the reader to Ref.~\cite{james} for more details.) We solve for the fluid velocity 
by using the standard pseudo-spectral method with  $N^3=512^3$ collocation points and a 
second-order Adams-Bashforth scheme to integrate in time. We drive the system to 
a statistically steady state by using a constant, large-scale energy injection forcing~\cite{pope-forc,pandit} one to reach 
the Taylor-scale Reynolds number $Re_{\lambda } \simeq 120$. 

To obtain the translational and orientation statistics, we seed the flow (as
obtained above) with (non-interacting) particles with seven different values of
$0.1 \le \alpha \le 2$ (including the spherical case $\alpha = 1$) and, including the tracers, six
different Stokes numbers $0.0 \le St_{\rm s} \le 3.0$; we use $N_p = 50000$ particles for
each $\alpha-St$ combination. We also run our simulations for several
large-eddy-turn-over times to rule out transient effects and obtain
well-converged statistics.  The trajectories of individual particles are
integrated by using a trilinear interpolation scheme ~\cite{NumRec} to obtain
the fluid velocity at the particle position. We set up an initial condition for the 
spheroids such that their orientation vector initially ($t=0$) points 
along the ${\bf \hat{x}}$ direction.


We begin by examining the alignment of the spheroids as a function of the
Stokes number and the aspect ratio. A convenient measure of the flow geometry
is to exploit the bases of the symmetric tensor ${\bf S}$ and the
anti-symmetric tensor ${\bf \Omega}$. Given the nature of the strain-rate
matrix, it is trivial to see that it allows three eigenvalues $\lambda_1 \ge
\lambda_2 \ge \lambda_3$ which correspond to a set of three orthonormal
eigenvector basis ${\bf S}{\bf e}_i = \lambda_i {\bf e}_i$.  The vorticity
tensor is constructed from the vorticity vector ${\bf \omega}$ yielding a
unit vector ${\bf e_\omega}$ corresponding to the magnitude of the vorticity
$\omega$.

\begin{figure*}
  \begin{center}
    \includegraphics[width=0.495\textwidth,trim=0 0 50 20,clip]{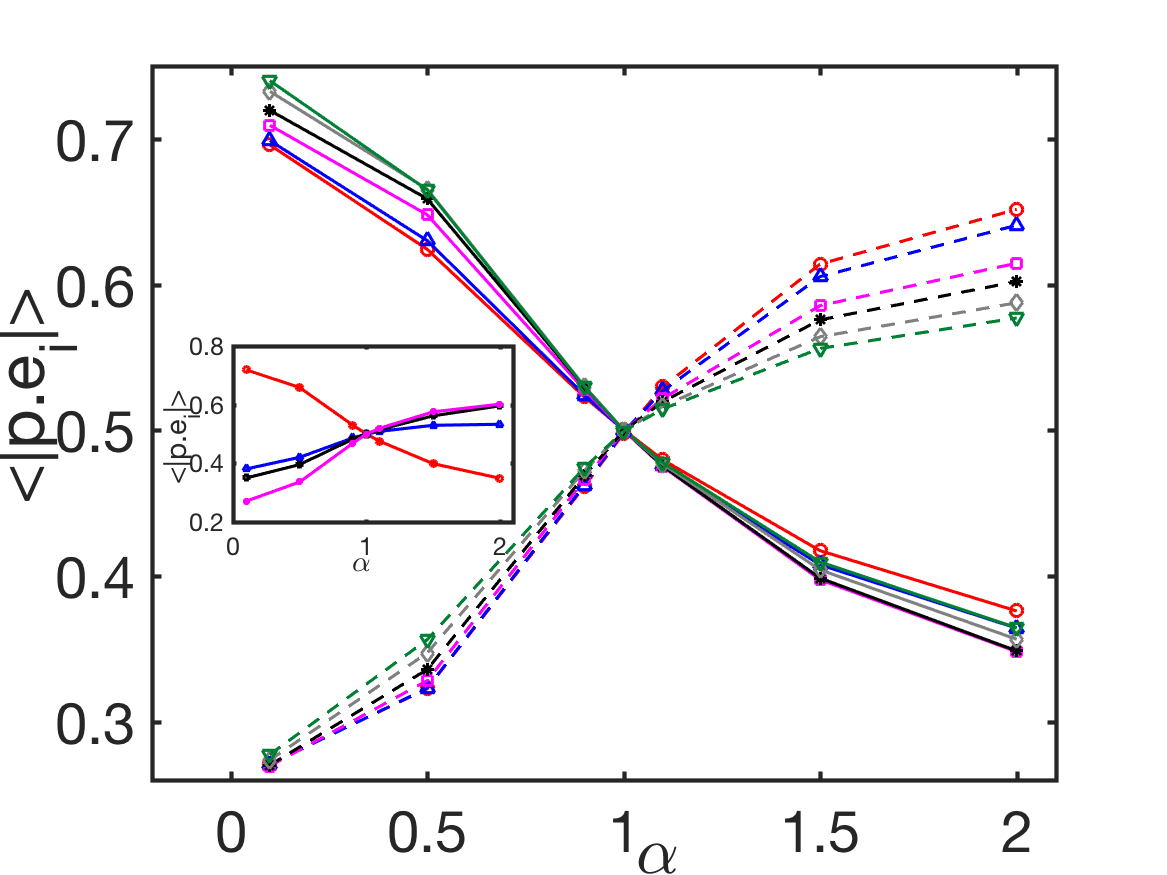}
\put( -40,171){\color{black}{\large (a)}}
    \includegraphics[width=0.495\textwidth,trim=0 0 50 20,clip]{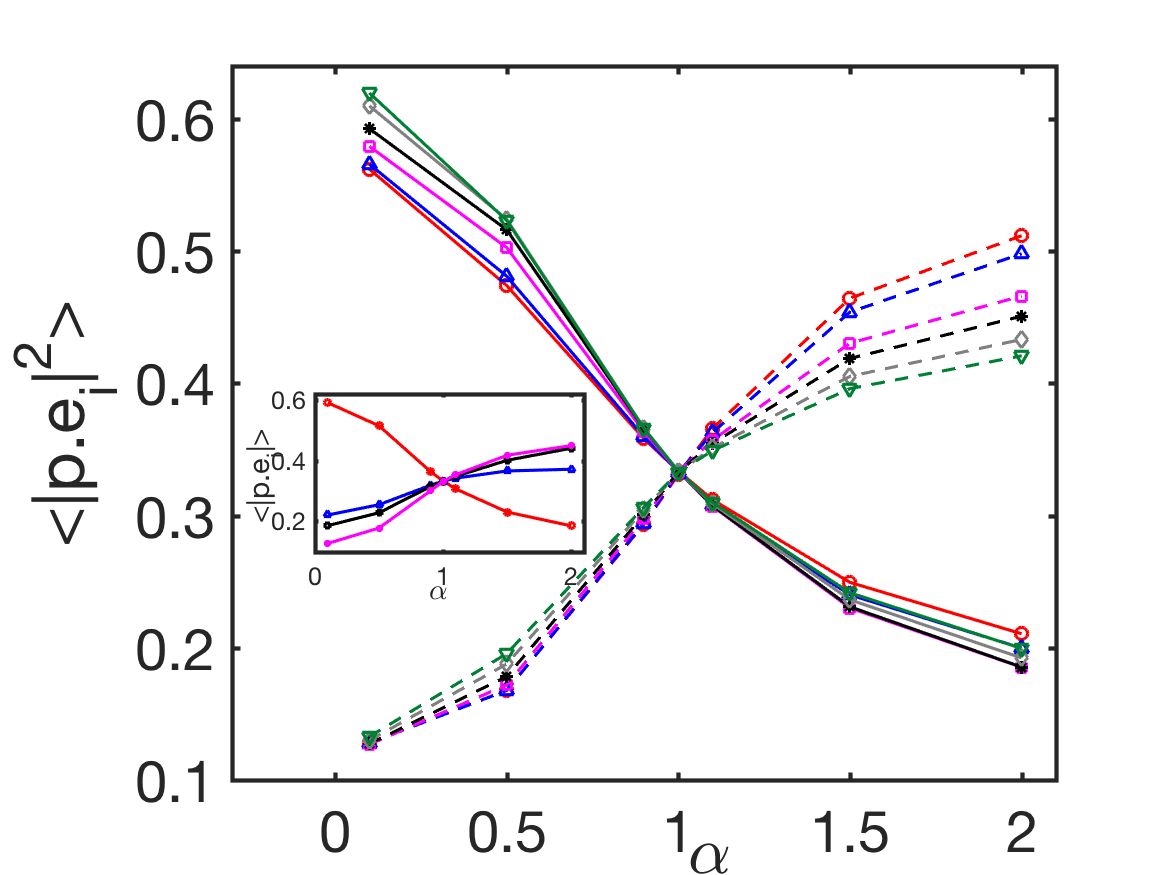}
\put( -40,171){\color{black}{\large (b)}}
    \vspace{-10pt}
\caption{(color online) Plots of (a) $\langle |{\bf p}.{\bf e}_i| \rangle$ and (b) $\langle |{\bf p}.{\bf e}_i|^2 \rangle$
vs $\alpha$ for $\Sts = 0$ (red open circles), $\Sts = 0.1$ (blue upward-pointing triangles), $\Sts = 0.5$ (magenta squares), 
$\Sts = 1.0$ (black asterisks), $\Sts = 1.5$ (gray diamonds), $\Sts = 2.0$ (green downward-pointing triangles). The solid and dashed 
lines are for ${\bf e}_1$  and ${\bf e}_\omega$, respectively.  
The insets show the representative plots of the same quantities, at $\Sts =1$, for ${\bf e}_1$ (red open circles),  
${\bf e}_2$ (blue triangles), ${\bf e}_3$ (black squares), and ${\bf e}_\omega$ (magenta asterisks).} 
\label{fig:mean}
  \end{center}
\end{figure*}

We characterize the alignment of the spheroids by calculating the probability
distribution function of the cosine of the angle between their orientation
vector with the different eigenvectors of the flow field.  The equation of
motion for the orientation vector suggests that $\bf p$ ought to align
preferentially with the principle axis of the strain rate matrix ${\bf e}_1$.
Surprisingly, however, it was shown by Pumir and
Wilkinson~\cite{wilkinson-NJP}, that measurements for tracers are inconsistent
with this na\"ive conclusion.  In Fig.~\ref{fig:alignment_vs_St}(a) we confirm
this conclusion from our numerical simulations. Given the plausible explanation 
for this phenomenon~\cite{wilkinson-NJP}, it is important to examine the effect 
of finite Stokes numbers. This is especially important because inertial spheroids 
will sample, preferentially, straining regions of the flow. 

In Fig.~\ref{fig:alignment}, we show representative plots of this probability
distribution function, namely $P(|{\bf p}.{\bf e}_i|)$ vs $|{\bf p}.{\bf
e}_i|$, where $i = 1, 2, 3$, and $\omega$, for different values of  $\alpha$
(for the same bare Stokes number of unity), calculated at times longer than the
initial transient phase (see Fig. 1 in Ref.~\cite{wilkinson-NJP}). Unlike the
tracer case, we see a very different behavior. For inertial oblates [Figs.~\ref{fig:alignment}
(a) and (b)], the spheroid tends to preferentially align with the principle
axis of the strain rate matrix as we should expect from the equation of motion
for the orientation vector. This behaviour is in contrast to rods ($\alpha >
1$) as shown in Figs.~\ref{fig:alignment} (c) and (d) where the alignment is most strongly with
the vorticity direction ${\bf e}_\omega$ 
as has been known for tracers~\cite{chevillardjfm}.
This behavior for rods is completely consistent with
what is known for tracer rods~\cite{wilkinson-NJP} and illustrated in
Fig.~\ref{fig:alignment_vs_St}(a). However unlike the $St = 0$ case, for finite
inertia rods tend to align to a greater degree with the non major axes of the
strain rate matrix, namely ${\bf e}_2$ and ${\bf e}_3$. Indeed this effect is enhanced 
for a given rod ($\alpha = 2.0$) with increasing inertia. In Fig.~\ref{fig:alignment_vs_St} 
we show representative plots of the probability density function for a rod 
with increasing values of the Stokes number from Figs.~\ref{fig:alignment_vs_St} (a) to (d). We clearly 
see that as the Stokes number increases, rods tend to align more and more with the 
axis ${\bf e}_3$ and, eventually, for the largest Stokes number considered here ($St = 4.56$, 
Fig~\ref{fig:alignment_vs_St}d), the alignment is strongest with ${\bf e}_3$ instead of ${\bf e}_\omega$ 
(Fig.~\ref{fig:alignment_vs_St}(a)).
For small inertia, rods tend to align with ${\bf e}_\omega$; however 
with increasing translational inertia, these spheroids start preferentially sampling 
strain-dominated regions. Hence, as the Stokes number increases, the rods start de-aligning 
with ${\bf e}_\omega$ and aligning with the most contracting eigenvector ${\bf e}_3$ (as 
clearly seen in our measurements) 
because the vorticity is normal to the most contracting direction~\cite{meneveauARFM}.

\begin{figure*}
  \begin{center}
    \includegraphics[width=0.495\textwidth,trim=0 0 50 20,clip]{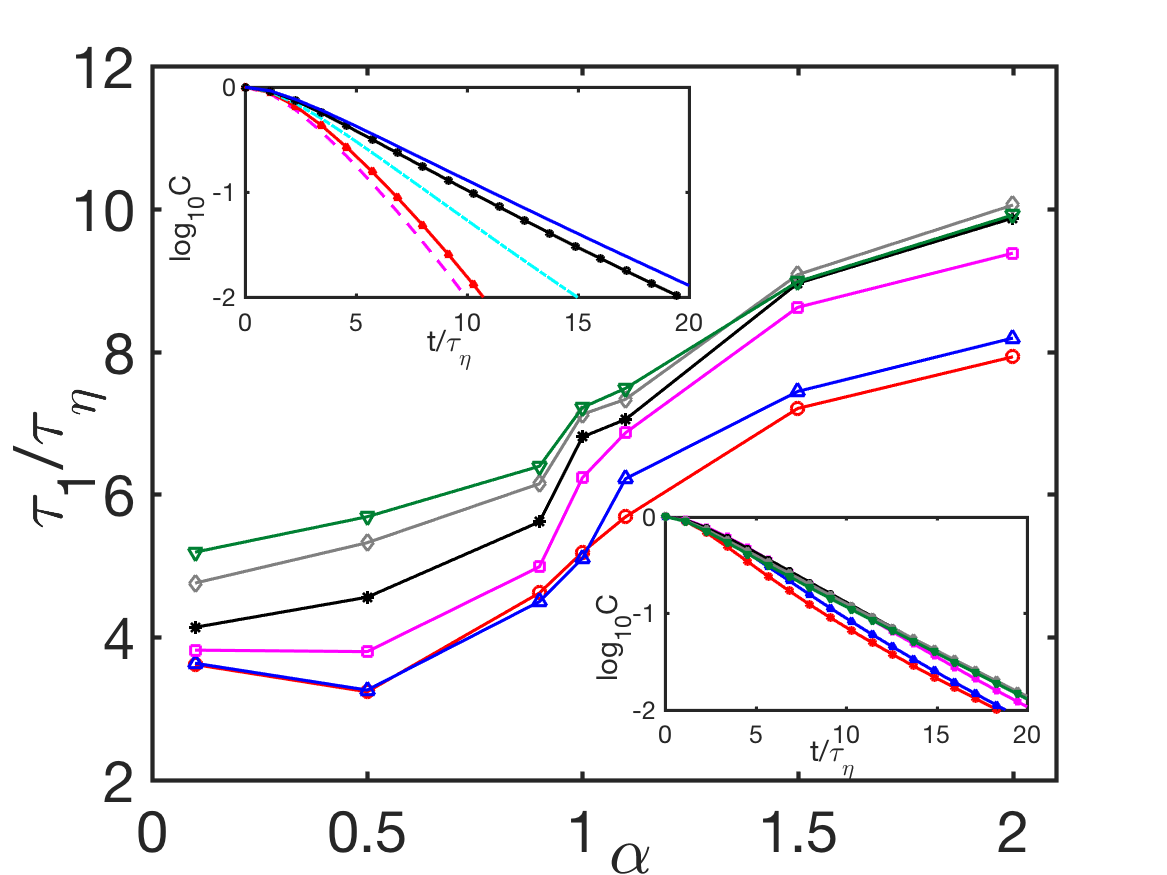}
\put( -40,171){\color{black}{\large (a)}}
    \includegraphics[width=0.495\textwidth,trim=0 0 50 20,clip]{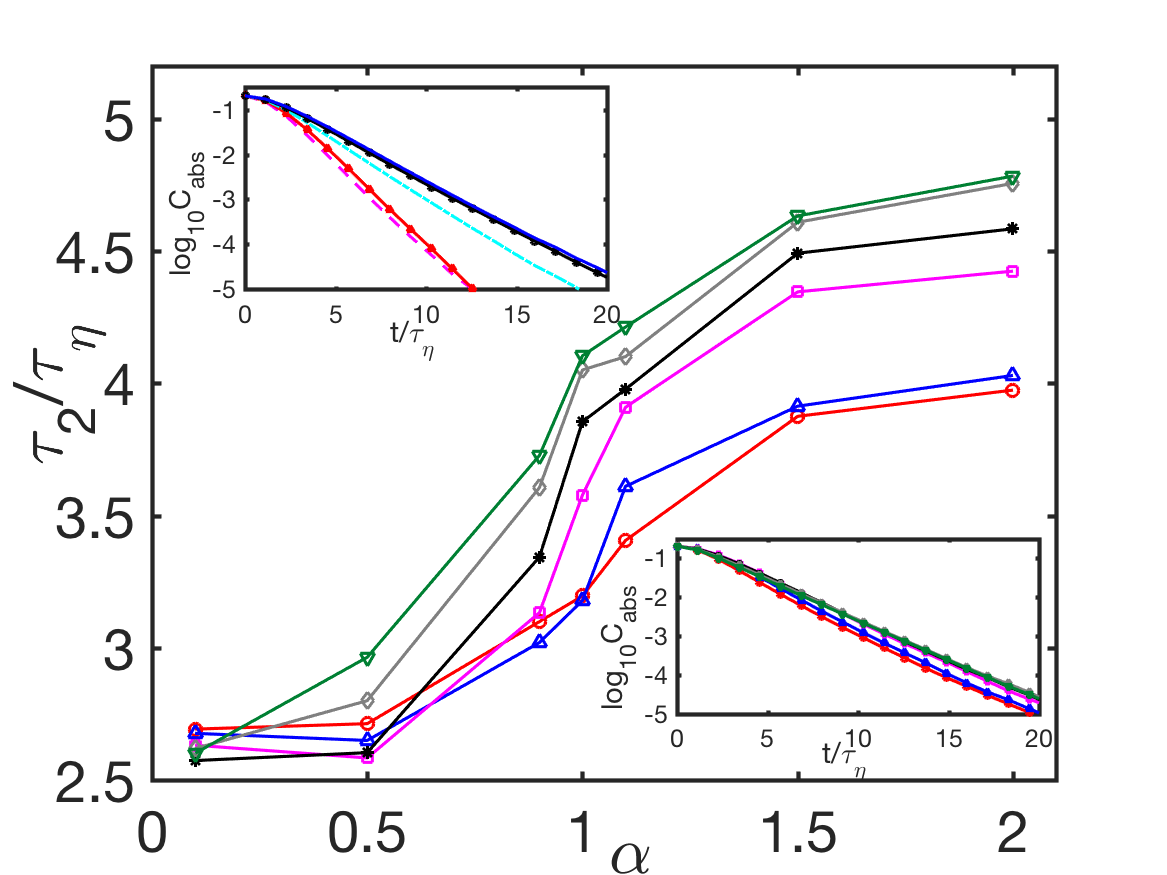}
\put( -40,171){\color{black}{\large (b)}}
    \vspace{-10pt}
    \caption{(color online) Plot of characteristic decay times, normalized by $\tau_\eta$, of the correlation functions (a) $\langle {\bf p}(t).{\bf p}(0) \rangle$ and (b) $\langle |{\bf p}(t).{\bf p}(0)| \rangle$  vs $\alpha$ for different $\Sts$. The color codes are the same as in Fig.~\ref{fig:mean}.
The top-left insets show the correlation function decay for a fixed $\Sts = 1.0$ and changing $\alpha$ from $\alpha = 0.1$ (magenta dashed line), $\alpha = 0.5$ (red triangles), $\alpha = 1.0$ (cyan dashed-dotted line), $\alpha = 1.5$ (black open circles) and $\alpha = 2.0$ (blue solid line).   
The bottom-right insets show the correlation function decay for a fixed $\alpha = 2.0$ and different $\Sts$ (same color code as the main plot). In the y axis of the insets of the left panel $C = \langle {\bf p}(t).{\bf p}(0) \rangle$ and the right panel $C_{\rm abs} = \langle |{\bf p}(t).{\bf p}(0)| \rangle - 0.5$.} 
\label{fig:decay-time}
  \end{center}
\end{figure*}

Our results suggest, unsurprisingly, that the dynamics of oblates, spheres, and
rods are qualitatively different from each other. Indeed for spherical
particles,  we expect that for all Stokes numbers, the orientation vector
should rotate randomly, yielding, on average $\langle|{\bf p}.{\bf
e}_i|\rangle$ = 0.5 and $\langle|{\bf p}.{\bf e}_i|^2\rangle$ = 0.33. This
reasoning breaks down in the case of spheroids; indeed in the limiting case of
tracer-rods ($St = 0$ and $\alpha \to \infty$), the actual values of these
measures are quite far from the spherical case~\cite{wilkinson-NJP}. In order
to systematically study the mean orientation of inertial spheroids, we measure
$\langle|{\bf p}.{\bf e}_i|\rangle$ and $\langle|{\bf p}.{\bf e}_i|^2\rangle$.
In Fig.~\ref{fig:mean}(a) and (b), we show plots of $\langle|{\bf p}.{\bf
e}_i|\rangle$ and $\langle|{\bf p}.{\bf e}_i|^2\rangle$, respectively, for
${\bf e}_1$ and ${\bf e}_\omega$, as a function of the aspect ratio $\alpha$
for a few representative values of the Stokes numbers. For both these measures,
the alignment with respect to the principle axis of the strain rate matrix is
close to 1 in the limit $\alpha \to 0$ and decreases monotonically and
approaches 0 as $\alpha \gg 1$. This behavior is exactly opposite to the mean
alignment with respect to the vorticity eigendirection where both these
measures increases monotonically with $\alpha$ and saturates, asymptotically, as
$\alpha \gg 1$. We  note that in the limiting spherical case $\alpha = 1$,
$\langle|{\bf p}.{\bf e}_1|\rangle$ = $\langle|{\bf p}.{\bf e}_\omega|\rangle$
= 0.5 and $\langle|{\bf p}.{\bf e}_1|^2\rangle$ = $\langle|{\bf p}.{\bf
e}_\omega|^2\rangle$ = 0.33 as suggested earlier. Furthermore, we  observe that
$\langle|{\bf p}.{\bf e}_\omega|\rangle$ and $\langle|{\bf p}.{\bf e}_\omega|^2 \rangle$
does not change with $\St$ for disks where as they decrease monotonically with
$\St$ for rods. On the other hand for the case ${\bf e}_1$ these 
measures increase monotonically  with $\St$ for disks; for the rods, however, 
this value first decreases with $\St$, reaches a minimum at $\St
= 0.5$, and then increases with $\St$. Finally, we note that the mean values 
for the alignment with ${\bf e}_2$ and ${\bf e}_3$ are following the same trend as ${\bf e}_1$ 
as shown in the insets of Fig~\ref{fig:mean}. 


Although it is still difficult in an experiment to accurately measure the
different eigenvectors along the Lagrangian trajectory of an spheroid -- as we
have done above -- a surrogate measurement is the autocorrelation functions $C
\equiv \langle ({\bf p}(t).{\bf p}(0)) \rangle$, $C_{\rm abs} \equiv \langle
|{\bf p}(t).{\bf p}(0)| \rangle$ and $C_2 \equiv \langle |{\bf p}(t).{\bf
p}(0)|^2 \rangle$ which decay exponentially at short times.  At long times,
these correlations asymptote to values close to 0, 0.5 and 0.33, respectively
as discussed above.  We measure such correlation functions and extract the
characteristic decay time scales $\tau_1$, $\tau_2$, and $\tau_3$ associated
with each of these correlation functions. In Fig.~\ref{fig:decay-time} we show
representative plots of $\tau_1$ [Fig.~\ref{fig:decay-time}(a)] and $\tau_2$ [Fig.~\ref{fig:decay-time}(b)],
normalized by the Kolmogorov time-scale $\tau_\eta$, as a function of the
aspect ratio $\alpha$ for a few representative values of the Stokes numbers.
These results are consistent  for the case of oblates studied (for similar
inertia and aspect ratios)by Jucha, \textit{et al.}~\cite{jucha} as well as
converging to the  rod- and tracer-limits reported in  Ref.~\cite{wilkinson-NJP}. 

Our measurements show a monotonic increase with 
the aspect ratio $\alpha$ with a mild, but non-trivial, dependence on the level of inertia.  
For $\alpha = 0.1$ (disks), $\tau_1$ increases monotonically with an increase in $\St$, but for the largest
simulated $\alpha = 2$, case first $\tau_1$ increases reaches a maximum at $\St
= 1$ and saturates with increasing $\St$.
We note that the maximum characteristic time for $\alpha = 2$ (rods) reaches at $\St = 1$, 
which corresponds to the case where maximum clustering starts to happen in turbulent flow.


For spherical particles, $\alpha = 1$, the second term on the right-hand-side of Eq.~(\ref{jeffery}) 
is absent by definition. Hence for spherical particles, the characteristic time scale 
is set by $\Omega$. However, assuming the term $\left ({\bf S \, p} - ({\bf p \cdot S \, p})\,{\bf p} \right )$
to be positive definite, a na\"ive interpretation of Eq.~\ref{jeffery} suggests that for $\alpha < 1$, the 
time scales for disks ought to be less than those for spheres; similarly for  $\alpha > 1$, the 
time scales for rods should be larger than those for spheres. This interpretation is 
consistent with the numerical results reported in Fig.~\ref{fig:decay-time}.
More pertinently, the statistics of alignment (discussed above) suggests that, for example, 
for disks, inertia leads to the orientation vector being orthogonal, preferentially, to the 
vorticity of the flows which lie in the plane of the disk, and hence, 
to a faster rotation of the orientation vector. Such an argument suggests that oblates 
rotate faster than rods resulting in a smaller decorrelation time for oblates than for rods. With 
increasing inertia, however, there is a preferential sampling of strain-dominated regions by particles of 
all shapes. Hence this leads, inevitably, to a smaller rotation rate and hence a larger decorrelation time. Indeed our 
measurements (Fig.~\ref{fig:decay-time}) show this to be the case. For the extremal values of $\alpha$, namely $\alpha = 0.1$ and 
$\alpha = 2.0$, the maximum values of $\St$ are 0.45 and 4.56, respectively. 
Hence we find that the decorrelation times for oblates are monotonically increasing in time with the Stokes number whereas for rods 
the saturation behavior is consistent with the fact that significant clustering starts to take place after $\St > 1$.
It is important to stress that these arguments are far from rigorous but 
seems to be consistent with our observations.

The rotational dynamics of small, but non-spherical, particles in turbulent flows is an important problem in many areas of 
fluid mechanics. In recent years, because of all the reasons mentioned earlier, there has been a lot of 
work in this area. However by and large most numerical and theoretical efforts have tended to ignore the 
effect of inertia -- and hence preferential sampling of the fluid velocity -- on the alignment properties of 
such particles. Furthermore even for the tracer case most studies have typically concentrated on the problem 
of rods. In this Rapid Communication, we have therefore systematically studied this problem by including the effects 
of inertia, for a large interval of aspect ratios spanning both disks and rods, to elucidate the statistics of 
the directional vector with respect to the geometry of the advecting flow. Our results show that the case of tracer rods, 
studied earlier, is a special case of spheroids and does not easily generalize for finite Stokes numbers or 
for disks. An important implication of our results lie in the modeling of asymmetrical microorganisms and the 
emergence of collective behavior (under suitable interactions) in a flow~\cite{unpublished}. 

\begin{acknowledgements}
SSR acknowledges the support of the DAE, Indo--French Center for Applied Mathematics (IFCAM) and the 
Airbus Group Corporate Foundation Chair in Mathematics of Complex Systems established in ICTS.
AR and SSR acknowledges the support of the DST (India) Project No. ECR/2015/000361. The simulations were performed 
on the cluster {\it Mowgli} and workstations {\it Goopy} and {\it Bagha} at the ICTS-TIFR. 

\end{acknowledgements}

\end{document}